# An experimental approach to mapping chemical bonds in nanostructured materials.

Philip N. H. Nakashima[1], Ding Peng[2], Xiaofen Tan[1], Anna N. Mortazavi[3,4], Tianyu Liu[1,5], Joanne Etheridge[1,5], Laure Bourgeois[1,5], David R. Clarke[3]

[1]Department of Materials Science and Engineering, Monash University, Victoria 3800, Australia.
[2]Department of Physics, Norwegian University of Science and Technology (NTNU), Trondheim, Norway.
[3]School of Engineering and Applied Sciences, Harvard University, Cambridge, USA.
[4]Department of Physics, Chalmers University of Technology, Gothenburg, Sweden.
[5]Monash Centre for Electron Microscopy, Monash University, Victoria 3800, Australia.

**Abstract**

We introduce a number of techniques in quantitative convergent-beam electron diffraction under development by our group and discuss the basis for measuring interatomic electrostatic potentials (and therefore also electron densities), localised at sub-nanometre scales, with sufficient accuracy and precision to map chemical bonds in and around nanostructures in nanostructured materials. This has never been possible as experimental measurements of bonding have always been restricted to homogeneous single-phased crystals.

**Introduction**

The electronic structure of chemical bonds is the dominant determinant of almost all materials properties[1] and it is in the accurate experimental measurement of chemical bonding as well as its modelling and interpretation by quantum theory, that the modern field of quantum crystallography resides.[2]

Originally the domain of X-ray diffraction experiments, quantum crystallography now also embraces quantitative convergent-beam electron diffraction (QCBED) measurements of quantum mechanical observables in crystals, namely the Fourier coefficients of the crystal potential (called structure factors).[2-4] Transforming crystal potential structure factors into electron density structure factors (measured directly by X-ray diffraction) is trivial via the Mott-Bethe formula[2, 4-6] and it is in the measurement of bonding-sensitive structure factors that QCBED has made a name for itself in the last few decades, in terms of both precision and accuracy.[2, 4, 7-56]

By definition, QCBED is the fitting of a calculated CBED pattern to an experimental one while adjusting the parameters to which the intensities in the pattern are most sensitive in order to minimise the pattern mismatch. Fig. 1 schematically illustrates CBED in the context of a specimen and gives an example of the comparison of experimental and calculated CBED intensities within QCBED.

A significant benefit of QCBED, besides accuracy and precision, is its spatial selectivity. It is routine to form focussed electron probes with nanometre or smaller dimensions in the process of obtaining convergent-beam electron diffraction (CBED) patterns. This is by virtue of electrons being charged and thus being easily manipulable by magnetic and electrostatic optical elements in electron microscopes. This means that CBED patterns can be acquired from volumes of crystal of order ~$10^9$ times smaller than in X-ray diffraction experiments, even at synchrotrons. Furthermore, these tiny electron probes can be positioned with sub-Ångström lateral precision to avoid inhomogeneities such as precipitate phases and crystal defects as shown in Fig. 1. Our new research concerning bonding in and around nanostructures in inhomogeneous materials means that we will be aiming to do the opposite of what is illustrated in Fig. 1: instead of targeting only the perfect regions of the matrix material, we will selectively probe the nanostructures as well. With the aid of computer-controlled scanning of the CBED probe over areas of, in principle, arbitrary user-defined shape, QCBED measurements of localised bonding structure as a function of position in nanostructured materials become a very interesting prospect.

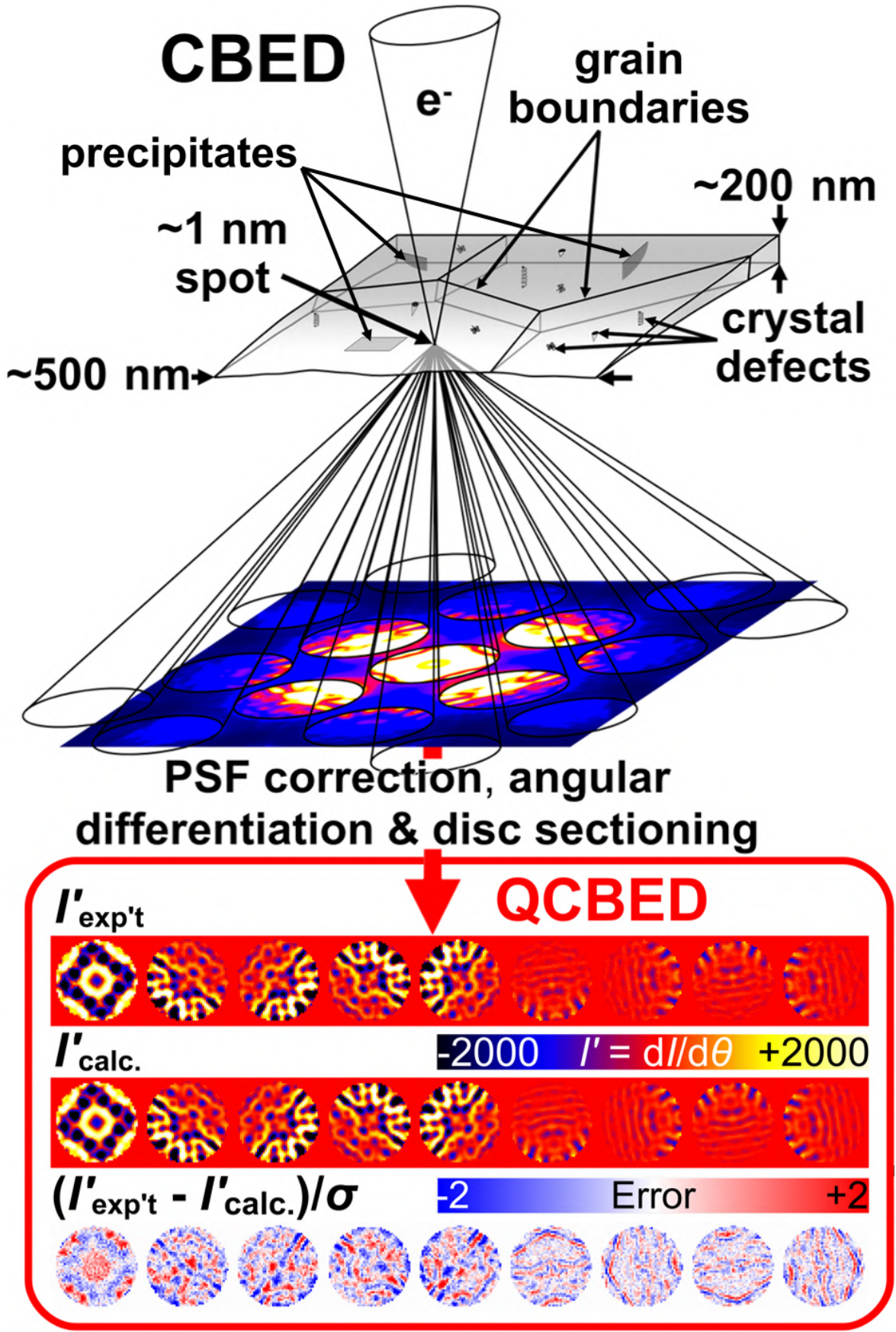

**Figure 1: A schematic illustration of CBED and pattern matching within QCBED.** QCBED involves the collection of an experimental CBED pattern from a real material, often replete with crystal defects, using a probe size of the order of 1 nm or smaller. The experimental intensities are pre-processed to eliminate errors due to instrumental point spread function (PSF) and inelastic scattering.[44, 57, 58] QCBED involves the fitting of a calculated CBED pattern to an experimental one while refining the parameters to which the intensities ($I$) – and therefore also the derivative of the intensities with respect to scattering angle ($I' = dI/d\theta$) – are most sensitive, in order to minimise the mismatch (shown here in multiples of the standard uncertainty of the experimental data, $\sigma$, in every pixel).

Nanostructures are, by definition, 3-dimensional of course and whilst sub-Ångström positioning of a nanometre or smaller sized electron probe in the *x* and *y* coordinates of a transmission electron microscope (TEM) has become routine, the ability to resolve structural information along the electron beam direction (defined as *z*) is a complex challenge. This difficulty is one that we are developing new QCBED techniques to overcome. We discuss these in this paper with the view to expanding bonding measurements from the current domain of homogeneous, single-phased crystals, to nanostructured materials.

Among the nanostructured materials that we are focussing on are aluminium alloys[59–65] and self-assembling thermoelectric $(ZnO)_k In_2O_3$ (k is an integer) superlattices.[66-69] These systems present situations in which there is no three-dimensional periodicity at all (aluminium alloys and some $(ZnO)_k In_2O_3$ thermoelectrics), or highly anisotropic three-dimensional periodicity where a unit cell of extreme aspect ratio can be defined (some thermoelectric $(ZnO)_k In_2O_3$ superlattices). In both types of materials, the layered structures within them are key to their desirable properties – aluminium alloys being ubiquitous in automotive, aerospace, packaging and structural materials applications, and thermoelectric $(ZnO)_k In_2O_3$ thin films being candidates for capturing and converting thermal energy normally lost in power plants, into electricity.

While the aim of this short paper is not to present or review results, we see it as an opportunity to discuss the measurement of bonding using electron diffraction in materials systems that have never before been accessible experimentally, namely nanostructured materials.

**Resolving atomic structure and bonding information along *z* – the beam direction**

Fig. 2 presents atomic-resolution high-angle annular dark field (HAADF) scanning transmission electron microscopy (STEM) images from our two families of materials that our future research will explore (Fig. 2a – d). Whilst the schematic illustration of a CBED probe at the same scale (Fig. 2e) grossly exaggerates the convergence angle (given that typical CBED convergence angles are of order 10 mrad), it serves to illustrate that the nanostructures to be probed are significantly greater in extent in directions perpendicular to the CBED probe than the dimensions of the probe itself, effectively rendering many nanostructures planar in the locality of the electron beam.

Fig. 2 also presents two CBED patterns. The first, Fig. 2f, was collected with the CBED probe aligned along the [001] direction in a $(ZnO)_k In_2O_3$ thermoelectric superlattice material, an example of which is imaged in the [1-10] direction in Fig. 2b. The second pattern was obtained with the probe incident in a near-[001] orientation on an Al-Cu-Sn alloy containing voids coated with single atomic layers of tin like the one shown in Fig. 2c and d.[65] The CBED probe was positioned to pass through the void, resulting in the pattern shown in Fig. 2g. The purpose of showing these diffraction patterns is twofold, namely: (i) to evidence that for nanostructures with crystallographically coherent interfaces throughout the probed volume and perpendicular to the electron beam direction, the CBED patterns have sharply defined reflection discs because the 2-dimensional projection of the reciprocal lattice along the beam direction is invariant as a function of depth within the specimen; and (ii) to show that there are many turning points in the intensity distributions within these patterns that will constrain the refinement of structural parameters, including the structure factors to which the pattern intensities are most sensitive – particularly those with short scattering vectors at or near their Bragg conditions (such as those that are shown in the present examples). These also contain the most bonding information.[9, 37, 53, 54]

Until recently, almost all QCBED in its modern incarnation was carried out by applying the Bloch-wave formalism for describing electron scattering.[7–30, 32–51, 53–56, 70] This theory assumes three-dimensional periodicity throughout the scattering volume and has restricted QCBED to homogeneous, single-phased crystalline materials.[2, 4, 7–56]

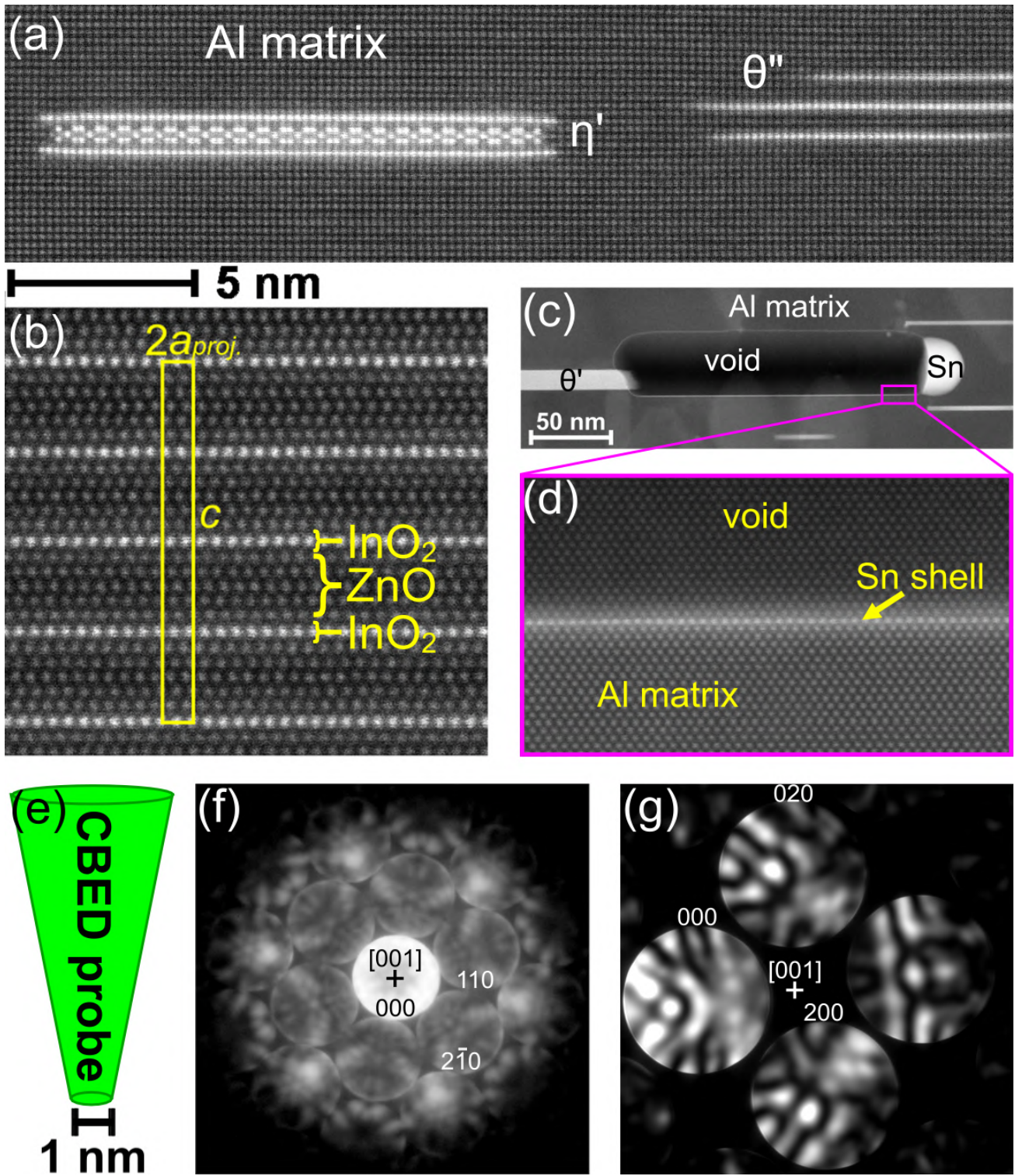

**Figure 2: Examples of nanostructures that are at least 2-dimensionally periodic in the context of a CBED probe and CBED patterns from two of these examples.** The first example (a) is a HAADF-STEM image of an aluminium alloy in a region containing two different precipitate phases that are both copper rich (brighter spots), namely the η' and θ" phases.[59–61, 63, 64] The second example (b) is a HAADF-STEM image from a thermoelectric $(ZnO)_kIn_2O_3$ superlattice thin film where k = 4.[66–69] The indium-containing layers, perpendicular to the plane of the image, manifest as brighter lines of spots. In this case, unlike the aluminium alloy, there is 3-dimensional periodicity and the projection of the unit cell has been drawn into the image. A more complex ensemble of nanostructures is shown in the third example (c) from an aluminium-copper-tin alloy containing voids attached to tin particles and θ' precipitates.[62, 65] The shell around each void in this system is composed of a single atomic layer of tin (d).[65] The high-resolution images (a, b and d) are at the same magnification as indicated by the scale bar common to these images, with the lower resolution HAADF-STEM image (c) having its own scale bar. The CBED probe has been drawn schematically (e) and is shown at the same scale as the high-resolution HAADF-STEM images (a, b and d), albeit with the angle of convergence grossly exaggerated to show that it is a convergent electron beam. CBED patterns from a $(ZnO)_kIn_2O_3$ superlattice thin film and through a void in the aluminium-copper-tin alloy are also shown (f and g respectively) and were collected with 200 keV electrons in both cases.

Recent work using the stacked Bloch-wave method has shown great promise in using CBED to resolve structural discontinuities within bulk materials.[71–75] Nonetheless, the ultimate spatial limit of Bloch-wave methods, including the stacking approach, is a unit cell. In contrast, the multislice theory of electron scattering requires the subdivision of a material into slices perpendicular to the beam and these slices can be taken to the limit of infinitesimal thickness.[76] In practical terms, it is often important to divide crystal potentials into slices with subatomic thicknesses to model electron scattering with sufficient accuracy. This results in the multislice formalism being intrinsically suited to situations in which there is no structural periodicity in the $z$ direction (along the electron beam), as would indeed be the case for any inhomogeneous nanostructured material in which bonding is to be measured and mapped.

A small number of papers from the present group have applied the multislice theory of electron scattering to QCBED.[31, 52, 77] However, when it came to accurate structure factor measurements, only homogeneous systems were studied in this way.[31, 52] Such a situation in the present context is schematically illustrated in Fig. 3a and b. It is worth noting that QCBED does in fact have its historical origins in the multislice formalism with the pioneering work of Goodman and Lehmpfuhl[78]; and Voss, Lehmpfuhl and Smith[79].

Work on localised surface plasmon resonances (LSPRs) in aluminium nanovoids, involving some of the members of the present group, used multislice-based QCBED to determine the depths of two voids that appeared to be immediately adjacent to one another but were in fact separated by a significant vertical distance in the beam direction, as revealed by QCBED.[77] In that application of QCBED, the multislice model incorporated a block of empty slices, similarly to the situation schematically illustrated in Fig. 3c, and the extent and depth of this vacant block were refined. Work is currently under way to test if structural details including bonding-sensitive structure factors and vacancy concentrations and their effects can be measured in the region of material surrounding nanovoids. In the present illustration (Fig. 3c), the additional presence of a single atomic layer of tin which coats nanovoids in Al-Cu-Sn alloys is also being interrogated and is the subject of ongoing work within our group (XT).

Voids present an interesting situation in physical optics in which the electron beam will Fresnel propagate across an extended block of vacuum, as illustrated in Fig. 3c. Early results have shown CBED patterns to be extremely sensitive to the depth and $z$-dimension of voids as a consequence of the entanglement inherent in the dynamical scattering of electrons as well as their propagation through free space.

Work has also commenced on layered $(ZnO)_k In_2O_3$ systems, with CBED data such as the pattern in Fig. 2f collected from regions of material such as the one shown in Fig. 2b and Fig. 3d. We are yet to collect CBED data from precipitates in aluminium alloys that can also be described as layered structures as shown in Fig. 2a and Fig. 3d.

Ultimately, the principles upon which measurements of bonding in and around nanostructures in inhomogeneous materials hinge are:

(i) The systems probed must be periodic in the two dimensions perpendicular to the direction of the electron beam.
(ii) The electrostatic potentials in each of the slices making up a multislice calculation of dynamical electron scattering in CBED simulations, must have periodicities that are coherent from slice to slice. This is necessary to preserve coherence in the two-dimensional reciprocal lattices projected along the beam direction that ultimately determine the geometries of the resultant diffraction patterns.

If these conditions can be met within the locality of a focussed electron beam, and if the probe size is large enough to sample more than one repeating unit of the 2-dimesionally periodic structure perpendicular to the electron beam direction, then, in principle at least, it may be possible to measure bonding in nanostructured materials and map bonds within and around nanostructures and across their interfaces with the parent matrix in which they are embedded.

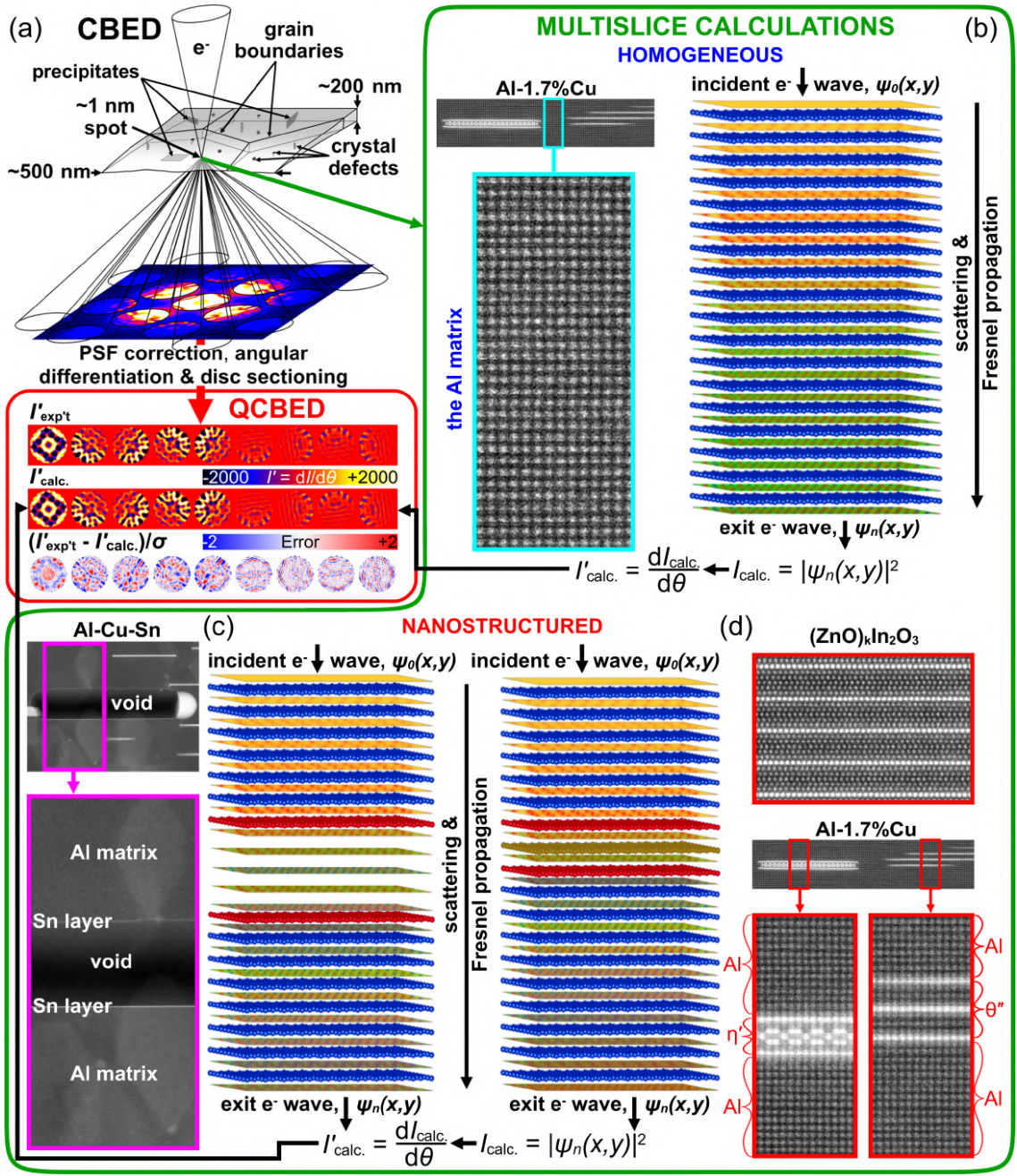

**Figure 3: Using the multislice formalism to perform QCBED of homogeneous and inhomogeneous materials.** Fig. 1 is repeated (a) in order to support the illustration of the different scenarios for applying QCBED in the context of nanostructured materials. Three scenarios are illustrated. In the first (b), the material volume to be probed (boxed in blue) is between two different precipitates in an Al-Cu binary alloy.[64] This region is homogeneous in three dimensions within the locality of a CBED probe. The corresponding multislice model is an uninterrupted sequence of planes of aluminium atoms (blue). The second case (c) is one in which the CBED probe is passed through a tin-coated nanovoid (the segment boxed in pink) in an Al-Cu-Sn alloy.[65] The corresponding multislice model involves the removal of a number of atomic planes – to create the void – plus the replacement of the bounding planes of aluminium atoms (blue) with planes of tin atoms (red) to model the tin coating. The third scenario (d) presents continuously solid regions of a $(ZnO)_k In_2O_3$ thermoelectric and $\eta'$ and $\theta''$ precipitates in an Al-Cu alloy.[64] The changes in atomic structure from layer to layer in these examples is accommodated by the introduction of atomic planes with different structures into the multislice model as is schematically illustrated by the red and yellow planes in the present figure (d). In all of the schematic illustrations of multislice models presented here, an attempt has been made to illustrate the modulation of the phase of the electron wave as it scatters from each slice of matter and propagates through the free space between slices. The exit wavefunction is used to calculate the scattered intensity making up the calculated CBED pattern being matched to the experimental one.

### *QCBED-DFT* and multislice

The field of X-ray constrained wavefunctions (XCW), well known in quantum crystallography, uses X-ray diffraction measurements to constrain many-body wavefunction calculations which return, among other things, the three-dimensional electron density distribution based on an exact quantum mechanical framework.[2, 80–85] This is analogous, in many ways, to *QCBED-DFT* – a new technique in which density functional theory (DFT) is embedded within QCBED so that DFT provides the crystal potentials used in the electron scattering calculations of CBED patterns instead of an independent atom model (IAM) as per conventional QCBED.[56] In *QCBED-DFT*, experimental CBED patterns constrain density functional and structural parameters that change the DFT-calculated electron distribution in real space. This in turn changes *all* structure factors in the Fourier sum that describes the electrostatic potential used in the calculation of the CBED patterns being matched to experimental ones (see Fig. 4), the latter being a direct consequence of the actual potential in the real material being probed in a TEM. As a consequence, *QCBED-DFT* is an answer to Walter Kohn's call in his Nobel Lecture, to constrain DFT with accurately known electron density distributions as opposed to the conventional approach of comparing DFT-calculated and experimentally measured materials properties and system energies.[86] It is worth noting that a three-dimensional electron density distribution is a much more complex (and therefore rigid) constraint than even very large sets of properties and energies.

Importantly, DFT is, in principle, exact, just like many-body wavefunction calculations. The use of functionals of the electron density makes DFT less computationally expensive than wavefunction methods as the number of atoms in a structure increases. However, for DFT to be exact in its calculations of both electron densities *and* system energies, it must be furnished with the exact density functional. The exact density functional remains one of the most hotly sought enigmas in science, let alone condensed matter physics.[86–91] Approximations abound in many different forms and are more or less suitable to different materials. We have already manifested our interest in *QCBED-DFT* as a means for testing different functionals as well as searching for the exact one, elsewhere.[56]

Here we express our interest in *QCBED-DFT* as a means for determining the bonding electron density in and around nanostructures within inhomogeneous materials. This interest is motivated by the nature of the synergy between QCBED and DFT, the key points being:

(i) DFT is constrained by experimental CBED data which are a direct consequence of the electron distributions within materials (as has already been pointed out).
(ii) The theoretical framework of DFT allows the interpretation of the *QCBED-DFT*-determined electron densities (and potentials) in terms of the large suite of materials properties that this theoretical framework makes accessible.[92-95]
(iii) A very complex parameter space can be compressed into a problem that has far fewer variables.

Having already discussed (i) above, and (ii) being a handy and self-contained statement of the obvious, our attention turns to point (iii), the basis of which derives from the fact that changing even a single, let alone a few, DFT model parameters, changes *all* structure factors and not just those to which the CBED pattern intensities are most sensitive, as explained above.

In any QCBED pattern-matching refinement, the fitting process will ultimately be driven by the sensitivity of the diffracted intensities to just a few structure factors, so in effect, within *QCBED-DFT*, those few structure factors are being refined by adjusting one or a few DFT model parameters that change all other structure factors according to the mould provided by the density functional being applied. If we now consider what this means in terms of the problems that we are setting out to solve in nanostructured materials, we can illustrate the implications via an example.

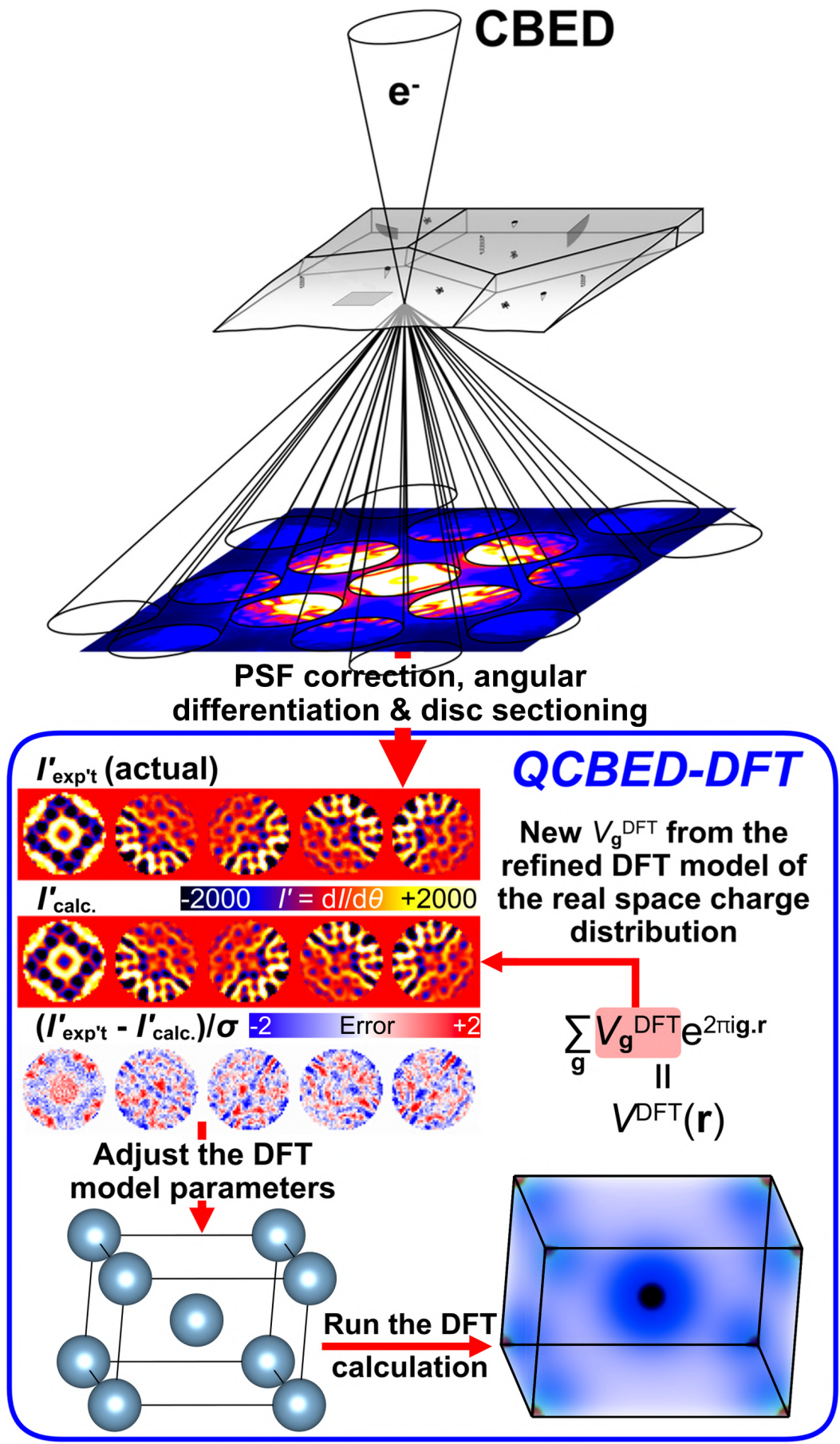

**Figure 4: *QCBED-DFT*.** This new technique builds density functional theory (DFT) into QCBED where previously, an independent atom model (IAM) was used to furnish the electron scattering calculations with the necessary crystal potentials. Instead of refining individual structure factors as per conventional QCBED, *QCBED-DFT* refines DFT model parameters which change the electron distribution in real space. This in turn changes *all* structure factors that are fed into the CBED pattern calculations.[56]

The periodic, two-dimensional sheets of $InO_2$, sandwiched between blocks of ZnO in $(ZnO)_kIn_2O_3$ thermoelectric materials are considered key to the properties of these superlattice thin-films since they act to scatter both phonons and conduction electrons.[66–69] It is therefore intuitive to ask about the nature of the chemical bonds in the $InO_2$ layers themselves as well as between these layers and their neighbouring ZnO layers. Setting out to do this by conventional QCBED could involve either the multislice or Bloch-wave formalisms in case of the example shown in Fig. 2b and Fig. 3d, owing to the three-dimensional periodicity, albeit with a very long *c*-axis unit cell (Fig. 2b). This would place quite severe requirements on the refinements of individual structure factors in order to be able to resolve bonding accurately within and around these layers of interest. Furthermore, a significant number of structure factors may be required to describe bonding at sufficient spatial resolution.

It is possible that *QCBED-DFT* may provide an avenue for simplifying this problem because rather than measuring many structure factors with the accuracy required to resolve bonding in and around the $InO_2$ planes, the structure factors will all be tied together by the density functional that generates them and that can only be manipulated and refined via a very small number of parameters. Here, we would be relying on the theoretical framework to fill a complex parameter space accurately, based entirely on the adjustments made to that framework via the tweaking of a small set of global parameters.

Having highlighted why *QCBED-DFT* has exciting potential for application to nanostructured materials, it is important to point out that the current form of *QCBED-DFT* requires three-dimensional periodicity because it is based on the Bloch-wave formalism for describing electron scattering.[70] This would work for the example of $(ZnO)_kIn_2O_3$ presented here in Fig. 2b and Fig. 3d (i.e. where k = 4) because of the three-dimensional periodicity that is present in this case, but it would fail for other $(ZnO)_kIn_2O_3$ systems where the spacings between $InO_2$ layers are non-uniform. It would also fail in the interrogation of the nanostructures in aluminium alloys that we have shown as examples in this paper (Fig. 2a, c and d; and Fig. 3c and d).

The ultimate method for measuring bonding in and around nanostructures within nanostructured materials would be *QCBED-DFT* based on the multislice theory of electron scattering and a priority of our future work will be the development of this approach. The software that will actuate it is in a beta stage of development by one of the authors (DP) at the time of writing. Ongoing programming will incorporate parallelisation methods introduced into recent iterations of our existing programs by another author (TL).

**Concluding remarks**

In this paper, we have presented no new results or reviewed any older ones but have instead discussed developments in QCBED that may pave the way to measuring bonding in and around nanostructures embedded in a parent matrix. Most materials in use today and in the future have and will have hybridised properties tailored to very specific applications where the target properties and combinations of different desired properties can only be achieved by the interaction between nanostructured precipitate phases and their surrounding host material. To understand these interactions and the properties that derive from them at the most fundamental level means that we must understand the nature of the chemical bonds within these nanostructures, within the host material and at the interfaces between them.

**Acknowledgements**

We thank the late Professor Andrew W. S. Johnson for being particularly influential in this area of research and for inspiring a number of its aspects. We are grateful to Professor Jian-Min Zuo for sharing his Bloch-wave QCBED code which has been modified by some of the authors over the last 3 decades to yield the QCBED methods at our disposal and that have been the subjects of this paper and are the cornerstones of the work being proposed. We thank Associate Professor Matthew Weyland for providing microscopy training, expertise and assistance in collecting some of the HAADF-STEM images presented in this paper. The authors acknowledge the use of instruments and the assistance rendered by the staff at the Monash Centre for Electron Microscopy, Monash University, a node of Microscopy Australia, established under the Commonwealth Government of Australia's National Collaborative Research Infrastructure Strategy (NCRIS). We are grateful to Mr Simon Michnowicz for allocating resources to this project from the Monash partner share of the National Computational Infrastructure (NCI). We acknowledge the assistance of resources and services from the NCI, which is supported by the Australian Government. This work has also been supported by the Monash e-Research Centre (MeRC), Monash University, via the MonARCH high-performance computing facility and we thank Mr Philip Chan, Mr Simon Michnowicz and Professor Wojtek Goscinski for their support and expertise. We thank Professor Randi Holmestad and Associate Professor Per Erik Vullum for supporting this work and DP acknowledges SFI Manufacturing, funded by the Research Council of Norway (grant number 237900) and several industry partners. We thank the Australian Research Council for funding (DP210100308).

**References**
[1]G. N. Lewis, *J. Am. Chem. Soc.* **38**, 762–785 (1916).
[2]A. Genoni, L. Bučinský, N. Claiser, J. Contreras-Garcia, B. Dittrich, P. M. Dominiak, E. Espinosa, C. Gatti, P. Giannozzi, J. -M. Gillet, D. Jayatilaka, P. Macchi, A. Ø. Madsen, L. J. Massa, C. F. Matta, K. M. Merz, P. N. H. Nakashima, H. Ott, U. Ryde, K. Schwarz, M. Sierka, S. Grabowsky, *Chem. Eur. J.* **24**, 10881–10905 (2018).
[3]L. Massa, L. Huang, J. Karle, *Int. J. Quant. Chem. Quant. Chem. Symp.* **29**, 371–384 (1995).
[4]P. N. H. Nakashima, *Struct. Chem.* **28**, 1319–1332 (2017).
[5]N. F. Mott, *Proc. R. Soc. A* **127**, 658–665 (1930)
[6]H. A. Bethe, *Ann. Phys. (N.Y.)* **5**, 325–400 (1930).
[7]J. M. Zuo, J. C. H. Spence, and M. O'Keeffe, *Phys. Rev. Lett.* **61**, 353–356 (1988).
[8]D. M. Bird and M. Saunders, *Ultramicroscopy* **45**, 241–251 (1992).
[9]J. C. H. Spence, *Acta Crystallogr. Sect. A* **49**, 231–260 (1993).
[10]J. M. Zuo, J. C. H. Spence, J. Downs, and J. Mayer, *Acta Crystallogr. Sect. A* **49**, 422–429 (1993).
[11]J. M. Zuo, *Acta Crystallogr. Sect. A* **49**, 429–435 (1993).
[12]C. Deininger, G. Necker, and J. Mayer, *Ultramicroscopy* **54**, 15–30 (1994).
[13]M. Saunders, D. M. Bird, N. J. Zaluzec, W. G. Burgess, A. R. Preston, and C. J. Humphreys, *Ultramicroscopy* **60**, 311–323 (1995).
[14]K. Tsuda and M. Tanaka, *Acta Crystallogr. Sect. A* **51**, 7–19 (1995).
[15]R. Holmestad, J. M. Zuo, J. C. H. Spence, R. Høier, and Z. Horita, *Philos. Mag. A* **72**, 579–601 (1995).
[16]J. M. Zuo and A. L. Weickenmeier, *Ultramicroscopy* **57**, 375–383 (1995).
[17]L. M. Peng and J. M. Zuo, *Ultramicroscopy* **57**, 1–9 (1995).
[18]M. Saunders, D. M. Bird, O. F. Holbrook, P. A. Midgley, and R. Vincent, *Ultramicroscopy* **65**, 45–52 (1996).
[19]P. A. Midgley and M. Saunders, *Contemp. Phys.* **37**, 441–456 (1996).
[20]C. Birkeland, R. Holmestad, K. Marthinsen, and R. Høier, *Ultramicroscopy* **66**, 89–99 (1996).
[21]J. M. Zuo, M. O'Keeffe, P. Rez, and J. C. H. Spence, *Phys. Rev. Lett.* **78**, 4777–4780 (1997).
[22]J. M. Zuo, M. Kim, M. O'Keeffe, and J. C. H. Spence, *Nature* **401**, 49–52 (1999).
[23]M. Saunders, A. G. Fox, and P. A. Midgley, *Acta Crystallogr. Sect. A* **55**, 471–479 (1999).
[24]M. Saunders, A. G. Fox, and P. A. Midgley, *Acta Crystallogr. Sect. A* **55**, 480–488 (1999).
[25]K. Tsuda and M. Tanaka, *Acta Crystallogr. Sect. A* **55**, 939–954 (1999).
[26]S. L. Dudarev, L. M. Peng, S. Y. Savrasov, and J. M. Zuo, *Phys. Rev. B* **61**, 2506–2512 (2000).
[27]V. A. Streltsov, P. N. H. Nakashima, and A. W. S. Johnson, *J. Phys. Chem. Solids* **62**, 2109–2117 (2001).
[28]K. Tsuda, Y. Ogata, K. Takagi, T. Hashimoto, and M. Tanaka, *Acta Crystallogr. Sect. A* **58**, 514–525 (2002).
[29]B. Jiang, J. M. Zuo, N. Jiang, M. O'Keeffe, and J. C. H. Spence, *Acta Crystallogr. Sect. A* **59**, 341–350 (2003).
[30]J. Friis, B. Jiang, J. C. H. Spence, and R. Holmestad, *Microsc. Microanal.* **9**, 379–389 (2003).
[31]V. A. Streltsov, P. N. H. Nakashima, and A. W. S. Johnson, *Microsc. Microanal.* **9**, 419–427 (2003).
[32]B. Jiang, J. M. Zuo, J. Friis, and J. C. H. Spence, *Microsc. Microanal.* **9**, 457–467 (2003).
[33]J. Friis, G. K. H. Madsen, F. K. Larsen, B. Jiang, K. Marthinsen, and R. Holmestad, *J. Chem. Phys.* **119**, 11359–11366 (2003).
[34]Y. Ogata, K. Tsuda, Y. Akishige, and M. Tanaka, *Acta Crystallogr. Sect. A* **60**, 525–531 (2004).
[35]B. Jiang, J. Friis, R. Holmestad, J. M. Zuo, M. O'Keeffe, and J. C. H. Spence, *Phys. Rev. B* **69**, 245110 (2004).
[36]J. Friis, B. Jiang, J. C. H. Spence, K. Marthinsen, and R. Holmestad, *Acta Crystallogr. Sect. A* **60**, 402–408 (2004).
[37]J. M. Zuo, *Rep. Prog. Phys.* **67**, 2053–2103 (2004).
[38]P. N. H. Nakashima, *J. Appl. Cryst.* **38**, 374–376 (2005).
[39]J. Friis, B. Jiang, K. Marthinsen, and R. Holmestad, *Acta Crystallogr. Sect. A* **61**, 223–230 (2005).
[40]P. N. H. Nakashima, *Phys. Rev. Lett.* **99**, 125506 (2007).
[41]K. Tsuda, D. Morikawa, Y. Watanabe, S. Ohtani, and T. Arima, *Phys. Rev. B* **81**, 180102(R) (2010).
[42]X. H. Sang, A. Kulovits, and J. M. K. Wiezorek, *Acta Crystallogr. Sect. A* **66**, 685–693 (2010).
[43]X. H. Sang, A. Kulovits, and J. M. K. Wiezorek, *Acta Crystallogr. Sect. A* **66**, 694–702 (2010).

[44]P. N. H. Nakashima and B. C. Muddle, *Phys. Rev. B* **81**, 115135 (2010).
[45]P. N. H. Nakashima, A. E. Smith, J. Etheridge, and B. C. Muddle, *Science* **331**, 1583–1586 (2011).
[46]P. A. Midgley, *Science* **331**, 1528–1529 (2011).
[47]X. H. Sang, A. Kulovits, and J. M. K. Wiezorek, *Acta Crystallogr. Sect. A* **67**, 229–239 (2011).
[48]X. H. Sang, A. Kulovits, G. F. Wang, and J. M. K. Wiezorek, *Philos. Mag.* **92**, 4408–4424 (2012).
[49]P. N. H. Nakashima, *Opt. Lett.* **37**, 1023–1025 (2012).
[50]X. H. Sang, A. Kulovits, G. F. Wang, and J. M. K. Wiezorek, *J. Chem. Phys.* **138**, 084504 (2013).
[51]X. H. Sang, A. Kulovits, and J. M. K. Wiezorek, *Ultramicroscopy* **126**, 48–59 (2013).
[52]D. Peng and P. N. H. Nakashima, *J. Appl. Cryst.* **50**, 602–611 (2017).
[53]P. N. H. Nakashima, *Struct. Chem.* **28**, 1319–1332 (2017).
[54]J. M. Zuo and J. C. H. Spence, *Advanced Transmission Electron Microscopy* (Springer, New York, 2017).
[55]D. Peng and P. N. H. Nakashima, *Acta Crystallogr. Sect. A* **75**, 489–500 (2019).
[56]D. Peng and P. N. H. Nakashima, *Phys. Rev. Lett.* **126**, 176402 (2021).
[57]P. N. H. Nakashima and A. W. S. Johnson, *Ultramicroscopy* **94**, 135–148 (2003).
[58]P. N. H. Nakashima and B. C. Muddle, *J. Appl. Cryst.* **43**, 280–284 (2010).
[59]V. Gerold, *Z. Metallkd.* **45**, 599–607 (1954).
[60]H. Yoshida, *Scr. Metall.* **22**, 947–951 (1988).
[61]S. C. Wang and M. J. Starink, *Int. Mater. Rev.* **50**, 193–215 (2005).
[62]L. Bourgeois, G. Bougaran, J. F. Nie, and B. C. Muddle, *Philos. Mag. Lett.* **90**, 819–829 (2010).
[63]P. N. H. Nakashima, "The Crystallography of Aluminum and its Alloys" in *The Encyclopedia of Aluminum and its Alloys*, ed. George E. Totten, Murat Tiryakioğlu and Olaf Kessler, 488–586 (Boca Raton: CRC Press,16 Nov 2018).
[64]L. Bourgeois, Y. Zhang, Z. Zhang, Y. Chen, and N. V. Medhekar, *Nature Commun.* **11**, 1–10 (2020).
[65]X. Tan, M. Weyland, Y. Chen, T. Williams, P. N. H. Nakashima, and L. Bourgeois, *Acta Mater.* **206**, 116594 (2021).
[66]X. Liang, M. Baram, and D. R. Clarke, *Appl. Phys. Lett.* **102**, 223903 (2013).
[67]X. Liang, *Curr. Nanosc.* **12**, 157–168 (2016).
[68]S. Margueron, J. Pokorny, S. Skiadopoulou, S. Kamba, X. Liang, and D. R. Clarke, *J. Appl. Phys.* **119**, 195013 (2016).
[69]X. Liang and D. R. Clarke, *J. Appl. Phys.* **124**, 025101 (2018).
[70]H. A. Bethe, *Ann. Phys. (Berlin)* **392**, 55–129 (1928).
[71]R. S. Pennington, C. Coll, S. Estradé, F. Peiró, and C, T, Koch, *Phys. Rev. B* **97**, 024112 (2018).
[72]R. S. Pennington and C. T. Koch, *Ultramicroscopy* **155**, 42–48 (2015).
[73]R. S. Pennington and C. T. Koch, *Ultramicroscopy* **148**, 105–114 (2015).
[74]R. S. Pennington, F. Wang, and C. T. Koch, *Ultramicroscopy* **141**, 32–37 (2014).
[75]R. S. Pennington, W. Van den Broek, and C. T. Koch, *Phys. Rev. B* **89**, 205409 (2014).
[76]J. M. Cowley and A. F. Moodie, *Acta Crystallogr.* **10**, 609–619 (1957).
[77]Y. Zhu, P. N. H. Nakashima, A. M. Funston, L. Bourgeois, and J. Etheridge, *ACS Nano* **11**, 11383–11392 (2017).
[78]P. Goodman and G. Lehmpfuhl, *Acta Crystallogr.* **22**, 14–24 (1967).
[79]R. Voss, G. Lehmpfuhl, and P. J. Smith, *Z. Naturforsch.* **35A**, 973–984 (1980).
[80]D. Jayatilaka, *Phys. Rev. Lett.* **80**, 798–801 (1998).
[81]D. Jayatilaka and D. J. Grimwood, *Acta Crystallogr. Sect. A* **57**, 76–86 (2001).
[82]D. J. Grimwood and D. Jayatilaka, *Acta Crystallogr. Sect. A* **57**, 87–100 (2001).
[83]L. Chęcińska, W. Morgenroth, C. Paulmann, D. Jayatilaka, and B. Dittrich, *Cryst. Eng. Comm.* **15**, 2084–2090 (2013).
[84]A. Genoni, L. H. R. Dos Santos, B. Meyer, and P. Macchi, *IUCrJ* **4**, 136–146 (2017).
[85]M. Woińska, D. Jayatilaka, B. Dittrich, R. Flaig, P. Luger, K. Woźniak, P. M. Dominiak, and S. Grabowsky, *Chem. Phys. Chem.* **18,** 3334–3351 (2017).
[86]W. Kohn, *Rev. Mod. Phys.* **71**, 1253–1266 (1999).
[87]R. O. Jones, *Rev. Mod. Phys.* **87**, 897–923 (2015).
[88]M. G. Medvedev, I. S. Bushmarinov, J. Sun, J. P. Perdew, and K. A. Lyssenko, *Science* **355**, 49–52 (2017).
[89]M. Korth, *Angew. Chem., Int. Ed. Engl.* **56**, 5396–5398 (2017).
[90]R. Car, *Nature Chem.* **8**, 820–821 (2016).
[91]J. Sun, R. C. Remsing, Y. Zhang, Z. Sun, A. Ruzsinszky, H. Peng, Z. Yang, A. Paul, U. Waghmare, X. Wu, M. L. Klein, and J. P. Perdew, *Nature Chem.* **8**, 831–836 (2016).
[92]P. Blaha, K. Schwarz, F. Tran, R. Laskowski, G. K. H. Madsen, and L. D. Marks, *J. Chem. Phys.* **152**, 074101 (2020).
[93]F. Tran, J. Stelzl, and P. Blaha, *J. Chem. Phys.* **144**, 204120 (2016).
[94]A. D. Becke, *J. Chem. Phys.* **140**, 18A301 (2014).
[95]K. Burke, *J. Chem. Phys.* **136**, 150901 (2012).